\documentclass[preprint,superscriptaddress]{revtex4}
\usepackage{graphicx}
\usepackage{dcolumn}
\usepackage{bm}
\usepackage{float}
\usepackage{hyperref}
\begin{document}
\title{Observation of Quantum Interference in Molecular Charge Transport}

\author{Constant M. Gu\'{e}don*}
\affiliation{Kamerlingh Onnes Laboratorium, Leiden University, Niels Bohrweg 2, 2333 CA Leiden, The Netherlands}

\author{Hennie Valkenier*}
\affiliation{Stratingh Institute for Chemistry and Zernike Institute for Advanced Materials, University of Groningen, Nijenborgh 4, 9747 AG Groningen, The
Netherlands}

\author{Troels Markussen}
\affiliation{Center for Atomic-scale Materials Design (CAMD), Department of Physics,
Technical University of Denmark, DK-2800 Kgs. Lyngby, Denmark}

\author{Kristian S. Thygesen}
\affiliation{Center for Atomic-scale Materials Design (CAMD), Department of Physics,
Technical University of Denmark, DK-2800 Kgs. Lyngby, Denmark}

\author{Jan C. Hummelen}
\affiliation{Stratingh Institute for Chemistry and Zernike Institute for Advanced Materials, University of Groningen, Nijenborgh 4, 9747 AG Groningen, The
Netherlands}

\author{Sense Jan van der Molen}
\affiliation{Kamerlingh Onnes Laboratorium, Leiden University, Niels Bohrweg 2, 2333 CA Leiden, The Netherlands}

\date{\today}
\maketitle
\textbf{As the dimensions of a conductor approach the nano-scale, quantum effects will begin to dominate its behavior. This entails the exciting
possibility of controlling the conductance of a device by direct manipulation of the electron wave function. Such control has been most clearly
demonstrated in mesoscopic semiconductor structures at low temperatures. Indeed, the Aharanov-Bohm effect\cite{Webb1985}, conductance quantization
\cite{Wees1988, Wharam1988} and universal conductance fluctuations\cite{Beenakker1991} are direct manifestations of the electron wave nature. However, an
extension of this concept to more practical temperatures has not been achieved so far. As molecules are nano-scale objects with typical energy level
spacings ($\sim$ eV) much larger than the thermal energy at 300 K ($\approx 25$ meV), they are natural candidates to enable such a
break-through~\cite{Sautet1988,Andrews2008,Stafford2007,Ernzerhof2005,Markussen2010,Markussen2010a,Solomon2010}. Fascinating phenomena
including giant magnetoresistance, Kondo effects and conductance switching, have previously been demonstrated at the molecular level\cite{Schmaus2011,
Smit2002, Kubatkin2003, Venkataraman2006, Osorio2007, Mishchenko2010, vanderMolen2010}. Here, we report direct
evidence for destructive quantum interference in charge transport through two-terminal molecular junctions at room temperature. Furthermore, we
show that the degree of interference can be controlled by simple chemical modifications of the molecule. Not only does this provide the
experimental demonstration of a new phenomenon in quantum charge transport, it also opens
the road for a new type of molecular devices based on chemical or electrostatic control of quantum interference.}\\

The wave nature of electrons is fundamental to our understanding of almost all of chemistry. In fact, the very existence of molecular orbitals is a direct result of spatial confinement of electron waves. This in turn leads to pronounced reactivity variation at different sites of molecules.
The electron wave character also plays a key role in
mesoscopic physics, which studies quantum phenomena in charge transport. For example, the conductance properties of mesoscopic ring structures at low
temperatures are dominated by quantum interference. If the partial waves through both branches of the ring add up destructively (constructively) a
suppression (enhancement) of the conductance is observed. For certain classes of molecular junctions, a similar effect is expected
\cite{Andrews2008,Stafford2007,Ernzerhof2005,Markussen2010,Markussen2010a,Solomon2010}. However, in that case the picture of interference
resulting from distinct spatial paths is no longer valid. Instead, interference in a molecule must be described in terms of electron propagation via paths
of orbitals, differing not only in space, but also in energy. Since the properties of molecular orbitals can be manipulated by chemical design, quantum
interference promises control over the conductance of molecular devices at the wave function level. In fact, conductance tuning over orders of magnitude at
ambient temperatures comes within reach. Although variations in charge transfer rates within donor-bridge-acceptor molecules can be explained in terms of
interference \cite{Patoux1997,Ricks2010}, only indirect indications for interference have been found in molecular conductance experiments
\cite{Mayor2003,Fracasso2011}. Here, we provide unambiguous evidence for destructive quantum interference in two-terminal molecular junctions.\\

To investigate the influence of quantum interference on molecular conductance properties, we synthesized five rigid $\pi$-conjugated molecular wires (see
Supplementary Methods). The first two molecules (AQ-MT and AQ-DT, left in Fig. 1a) contain an anthraquinone-unit. This makes them cross-conjugated
\footnote{Linear conjugation refers to a sequence of alternating single and double bonds between both ends of an organic molecule. Cross-conjugation
implies that the sequence of alternating single and double bonds between both ends of the molecule is broken, although all C-atoms have formed double or
triple bonds, i.e. all C-atoms are sp$^2$ or sp hybridized.} \cite{Dijk2006, Gholami2006}. The AQ-MT molecule is terminated by a protected thiol
group at one side only (monothiolated: MT), whereas AQ-DT is dithiolated (DT). The third molecular wire (AC-DT) contains a central anthracene-unit and is
linearly conjugated. Otherwise it is very similar to AQ-DT, e.g. both have a length of 24.5 \AA. Finally, two linearly conjugated reference compounds,
oligo(phenylene-ethynylene)-monothiol and -dithiol (OPE3-MT and OPE3-DT), are studied. We stress that apart from the thiols,
all five molecules have the same phenylene-ethynylene endgroups.\\
To measure transport, we first create self-assembled monolayers (SAMs) of each molecule on thin Au layers (200 nm, Si-substrates). To obtain high-quality,
densely packed SAMs, we use a procedure established recently (Supplementary Methods) \cite{Valkenier2011}. Next, a conducting atomic force microscopy (AFM)
probe is brought in close contact to a SAM. In this way, we can perform charge transport experiments through the molecular layer, using the Au-covered
substrate and the AFM-tip as electrodes (Fig. 1b). We typically connect to a few hundred molecules, while measuring current, $I$, versus bias voltage $V$
\cite{Wold2001}. However, the exact number does vary. For this reason, we present our results in two-dimensional (2D) histograms. Figure 1c shows such a
2D-histogram for AC-DT. To construct this plot, we have logarithmically binned the ${dI}/{dV}$-values (determined numerically) for each bias applied (see
Supplementary Methods). This effectively results in a sequence of 1D-histograms, plotted for each $V$. To illustrate this, Fig. 1d shows a cross-section of
Fig. 1c at $V=0$ V (blue histogram; see dashed line in Fig. 1c). This is the zero-bias 1D-histogram for AC-DT \cite{Gonzalez2006}. Representing our data in
2D-histograms has two major advantages. First, it allows us to show a full data set in one plot, without a need for either determining an average curve or
for data selection \cite{Gonzalez2006} \footnote{For completeness: I(V)-curves that were either flat (no contact) or that showed direct contact
are excluded from Figs. 1 and 3. However, such curves represent a small minority of our data ($\approx 5 \%$), see Supplementary Methods.}. Second, it
enables us to distinguish general tendencies in $dI/dV$-curves from statistical variations in the conductance values themselves. The latter are inherent to
molecular transport studies \cite{Wold2001,Gonzalez2006}. Figure 1c clearly illustrates this advantage: a symmetric valley-like shape is seen for the full
data range. This shape is virtually independent of the conductance values, which do scatter indeed (Fig. 1d).\\

\begin{figure}
\centering
\includegraphics[width=1\textwidth]{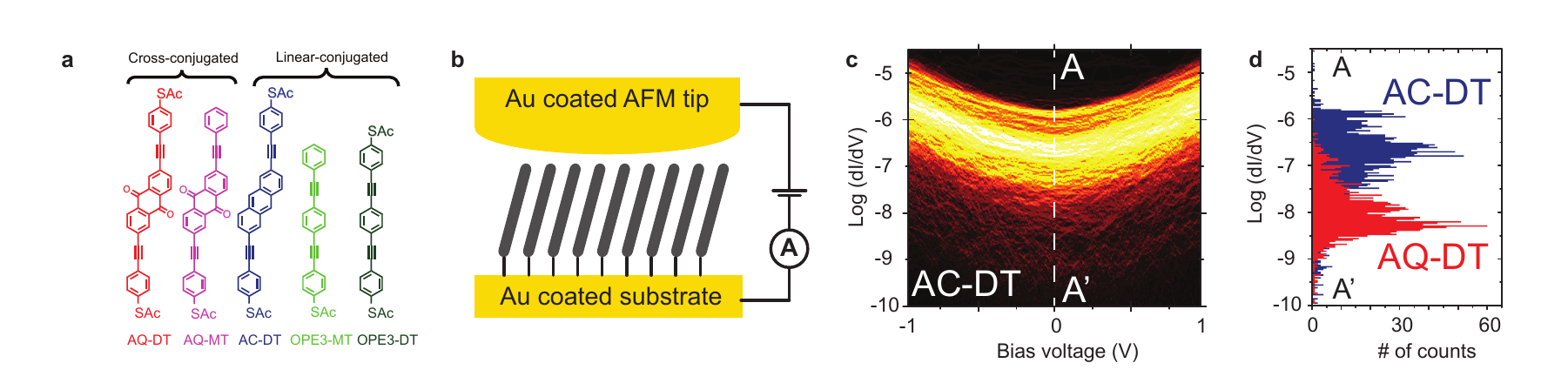}
\caption{\textbf{Conductance measurements on molecular wires}. \textbf{a}, chemical structure of the molecules used. AQ-DT and AQ-MT are both
cross-conjugated, whereas AC-DT, OPE3-DT and OPE3-MT are linearly conjugated. AQ-DT, AC-DT and OPE3-DT are dithiolated and thus symmetric; AQ-MT and
OPE3-MT are monothiolated. The colour code is also used in the following figures. \textbf{b}, schematic view of the junction formed by the molecules
self-assembled on a conducting substrate (Au) and the conducting AFM tip (Au), \textbf{c}, logarithmically binned 2D-histogram for the ${dI}/{dV}$-values
vs. bias voltage $V$ for AC-DT in $\Omega^{-1}$, the colour scale indicates the number of counts (black: no counts; white: more than 40 counts)
\textbf{d}, cross-section of the 2D-histogram shown in \textbf{c} along the line AA' (zero-bias conductance) resulting in a 1D-histogram (blue). Shown in
red is the 1D-histogram for AQ-MT taken from Fig. 3a}
\end{figure}

Figure 1d compares the zero-bias conductance histograms for both AQ-DT (red) and AC-DT (blue). Interestingly, AQ-DT exhibits conductance values that are
almost two orders of magnitude lower than those of AC-DT. This is quite remarkable, since the energy difference between the HOMO and LUMO levels is very
similar for these molecules (HOMO: highest occupied molecular orbital; LUMO: lowest unoccupied molecular orbital) \footnote{From UV-Vis measurements, we
find an optical HOMO-LUMO gap of 2.88 eV for AQ-DT and 2.90 eV for AC-DT. Our calculations yield fundamental HOMO-LUMO gaps of 4.23 eV and 4.61 eV,
respectively. Note that the optical gap and the fundamental gap differ by the electron-hole interaction.}. Furthermore, Fig. 1d cannot be trivially
explained from a weaker coupling of AQ-DT to the AFM-tip, since the endgroups of both molecules are exactly the same. As we shall elaborate on
below, the large difference in conductance is instead indicative of destructive interference in the AQ-DT junctions. In Fig. 2a we present calculations of
the energy-dependent transmission function, $T(E)$, for junctions containing AC-DT, AQ-DT, and AQ-MT. This function describes the quantum mechanical
probability that an electron with energy $E$ traverses the molecular junction. Once $T(E)$ is known, the $I(V)$-curves can be calculated using Landauer's
formula (Supplementary Methods). In particular, the low bias conductance is given by $dI/dV(V=0)=2e^2/h \cdot T(E=E_F)$. For a molecular junction, $T(E)$
typically exhibits peaks around the orbital energies, where electrons can tunnel resonantly. In the energy gaps, $T(E)$ is normally rather featureless, as
exemplified by AC-DT in Fig. 2a. However, for AQ-DT and AQ-MT, $T(E)$ exhibits a strong dip or 'anti-resonance'. This feature is a result of destructive
interference \cite{Andrews2008,Stafford2007,Ernzerhof2005,Markussen2010,Markussen2010a,Solomon2010}. To reveal the origin of the
anti-resonance, we transform the frontier molecular orbitals into an equivalent set of maximally localized molecular orbitals (LMOs)\cite{Markussen2010}.
The upper part of Fig. 2d shows the three relevant LMOs obtained for AQ-DT. Two are localized on the left and right parts of AQ-DT, respectively. These
LMOs have the same energy and correspond essentially to the sum and difference of the almost degenerate HOMO and HOMO-1 (Fig. 2a). The LMO localized in the
center of AQ-DT is essentially the LUMO and has a higher energy.  It is now clear that an electron with an energy, $E$, lying inside the HOMO-LUMO gap can
traverse the molecule via two distinct paths: either directly from the left to the right LMO or by going via the energetically higher LMO (arrows in Fig.
2d). It can be shown that the upper and lower routes yield a phase difference of $\pi$ within the HOMO-LUMO gap, i.e., the partial waves interfere
destructively (Supplementary Methods). Consequently, $T(E)$ shows a strong minimum at the energy where the partial waves have equal weight. Figure 2c
illustrates this, by showing $T(E)$ calculated for the lower and upper routes separately, as well as for the combined three-site model. Note the similarity
to Fig. 2a. For AC-DT, the HOMO is well separated from the HOMO-1. Hence, a transformation to LMOs leads to only two, left and right localized, orbitals
(Fig. 2b). As there is only a single path available, no interference effects occur for AC-DT.

\begin{figure}
\centering
\includegraphics[width=1\textwidth]{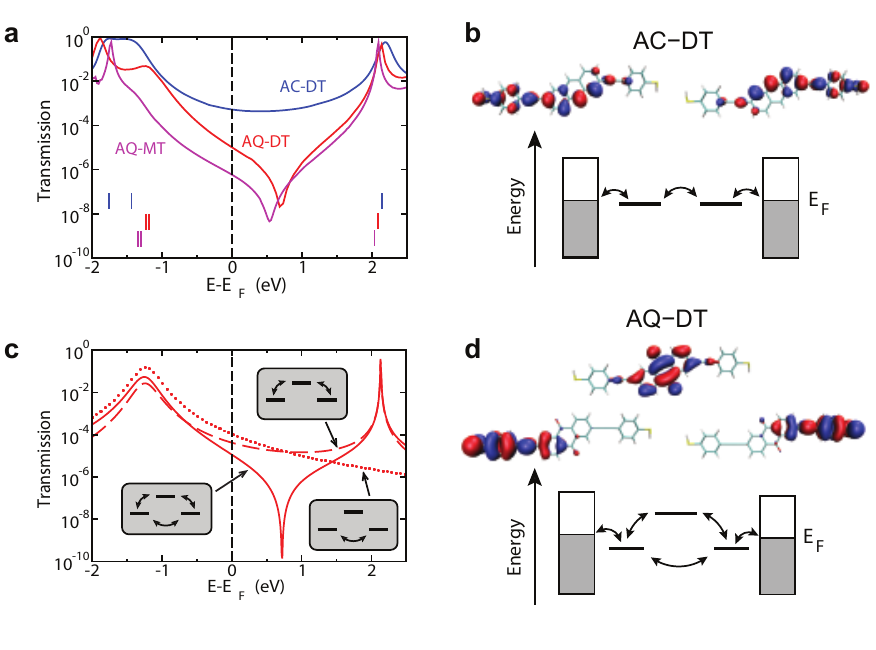}
\caption{\textbf{Origin of interference in cross-conjugated molecules}. \textbf{a}, Transmission functions $T(E)$ for AC-DT (blue), AQ-DT
(red) and AQ-MT (purple) calculated with DFT+$\Sigma$. The vertical bars mark the energies of the frontier orbitals HOMO-1, HOMO (left side) and LUMO
(right side). The lower part
of \textbf{b} and \textbf{d} pictures schematic transport models derived from the localized molecular orbitals presented in the upper parts. In the
three-site model shown in \textbf{d}, there are evidently two routes through the molecule: a lower route directly from the left to the right site and an
upper route via the central orbital. Panel \textbf{c} shows $T(E)$ for the lower (dotted line) and upper route (dashed line). A coherent addition of the
transmission probability amplitudes from the two paths, with a phase difference of $\pi$, yields the three-site transmission function (solid line).
This reproduces the essential features of \textbf{a}, for AQ-DT and AQ-MT.}
\end{figure}

We now compare these calculations with the experiments in Fig. 1d. In Fig. 2a, the $T(E_F)$-values are around two orders of magnitude lower for
AQ-DT than for AC-DT. This is in reasonable agreement with the strongly reduced conductance values for AQ-DT in Fig. 1d. We thus have a first indication of
interference in AQ-DT. To investigate this further, we inspect the full 2D-histogram of AQ-DT (Fig. 3a). For the full voltage range, its $dI/dV$-values are
dramatically lower than those of AC-DT (Fig. 1c). However, the 2D-histogram has a parabola-like appearance similar to AC-DT, i.e. we observe no anomaly
that can be connected to the calculated transmission dip. Hence, although the surprisingly low conductance of AQ-DT is most likely due
to quantum interference, the evidence is only indirect. This situation is comparable to the one in Refs. \cite{Mayor2003,Fracasso2011}\\
Let us next consider AQ-MT molecules, which should also exhibit an anti-resonance (Fig. 2a). Figure 3b shows the 2D-histogram of the $dI/dV$-curves for
AQ-MT (Supplementary Figures). Remarkably, these data do show a clear anomaly at zero bias voltage. In particular, the curvature of the $dI/dV$-traces in
Fig. 3b is negative for all $V$ (except around $V=0$). What is equally striking in Fig. 3b is the large voltage range over which the anomaly extends. Even
up to $V=\pm 1$ V, the $dI/dV$-curves are dominated by the minimum at $V=0$ V. This points to a characteristic energy scale of $\approx 1$ eV, which
corresponds well with the width of the interference-induced dip in $T(E)$ in Fig. 2a. Moreover, this large energy scale rules out Kondo effects and Coulomb
blockade as possible explanations for the anomaly \footnote{Coulomb blockade can also be ruled out via the experimental data. If Coulomb blockade were the
dominant effect behind Fig. 3b, it should also be present in the other molecular junctions, which have the same length and hence lead to roughly the
same capacitance. However, no anomaly is seen in Figs. 1c and 3a, 3c and 3d.}. Hence, Fig. 3b makes a strong case for quantum interference.\\

\begin{figure}
\centering
\includegraphics[width=1\textwidth]{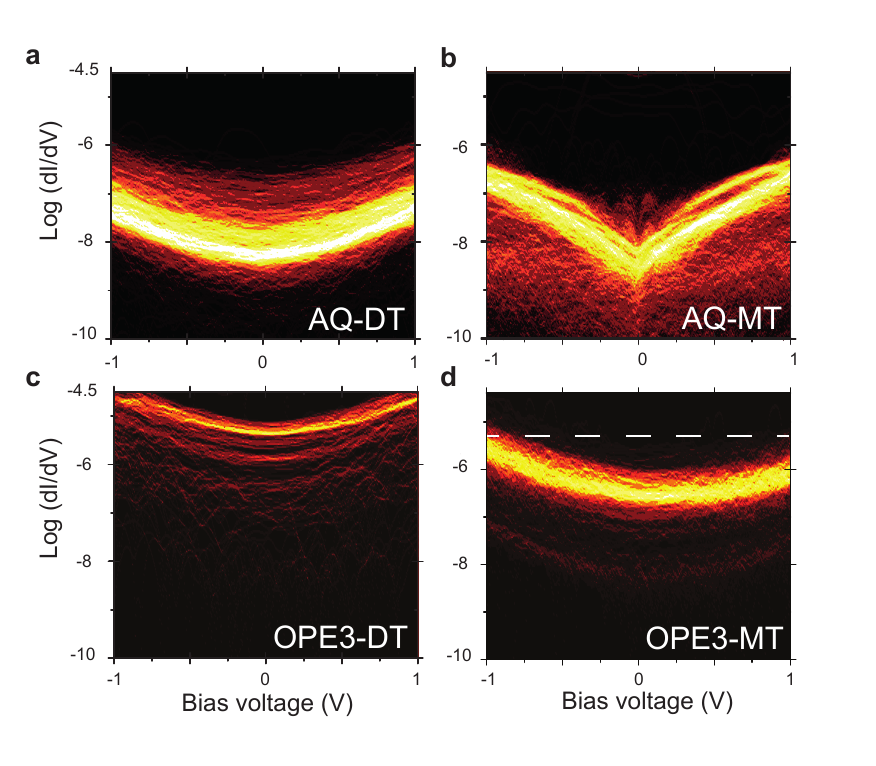}
\caption{\textbf{Two-dimensional conductance histograms}. \textbf{a-d} show logarithmically binned 2D-histograms of $dI/dV$ (in $\Omega^{-1}$)
vs. bias voltage $V$ for AQ-DT (\textbf{a}), AQ-MT (\textbf{b}), OPE3-DT (\textbf{c}) and OPE3-MT (\textbf{d}). The colour scale indicates the number of
counts and ranges from black (0 counts) to white (more than 40 counts). In \textbf{d}, a dashed line visualizes the asymmetry in the $dI/dV$-curves of
OPE3-MT, resulting from asymmetric coupling. The corresponding molecular structures can be found in Fig. 1a.}
\end{figure}

To further validate this interpretation, we calculate $dI/dV$-curves for AQ-MT from $T(E)$ (see Supplementary Methods). A key role in these
calculations is played by the position of the anti-resonance in $T(E)$ relative to $E_F$. This position is difficult to predict theoretically. This is
related to the well-known problems of the applied methodology to describe energy level alignments and to the uncertainty of the size of the surface dipoles
in the experiments \footnote{At Au-S interfaces, charge is transferred from Au to S, thus creating a surface dipole that shifts the molecular levels upward
in energy. This shift depends, among other factors, on the surface density of molecules.}. The position of the anti-resonance is particularly sensitive to
such effects due to the low density of states in the HOMO-LUMO gap \footnote{The computational limitation is illustrated best by comparing our calculations
on AQ-DT (Fig. 2) with those in Ref. \cite{Fracasso2011} (Fig. 5). In our Fig. 2, the anti-resonance lies to the right of $E_F$, whereas in Ref.
\cite{Fracasso2011}, it lies to the left.}. It is thus reasonable to treat the position of the transmission minimum as a free variable within a limited
energy window. In Fig. 4a, we display $dI/dV$-curves for AQ-MT, calculated for three cases: no energy shift (compared to Fig. 2a) and shifts of $\pm 0.5$
eV. We take into account that AQ-MT molecules are probed asymmetrically. For a shift of $-0.5$ eV, the calculated $dI/dV$-characteristic is in remarkable
agreement with the measured curves in Fig. 3b. First, the V-like shape with negative curvature is reproduced. Second, the voltage scale and the range of
$dI/dV$-values over which the minimum extends agree. Finally, the $dI/dV$-curves are nearly symmetric in both calculation and experiment. The latter is
indeed noteworthy, since AQ-MT is contacted asymmetrically. The symmetry in Fig. 3b must therefore be a consequence of $T(E)$ being symmetric around $E_F$
or, equivalently, of $E_F$ laying near the interference minimum. To independently confirm that monothiols are asymmetrically coupled, we measured
$dI/dV$-curves for OPE3-DT (Fig. 3c) and OPE3-MT (Fig. 3d). As expected, symmetric data are obtained for OPE3-DT, whereas asymmetric $dI/dV$-curves are
found for OPE3-MT (see Fig. 4b for calculations). Hence, we conclude that Fig. 3b constitutes direct proof for quantum interference in AQ-MT molecular
junctions \footnote{There is still the question why AQ-DT does not show a V-shaped $dI/dV$-curve, while its conductance is strongly suppressed.
This is explained by the fact that AQ-DT junctions comprise two Au-S dipoles, whereas AQ-MT junctions have only one. Hence, in AQ-DT, the transmission dip
is shifted to higher energies than in AQ-MT, i.e. it lies above $E_F$. In that case, no anomaly shows up in $dI/dV$-curves (see Figs. 4a and
S12). Note that a higher energetic position of the dip of AQ-DT, compared to AQ-MT, is also predicted by our calculations (Fig. 2a).}.\\

\begin{figure}
\centering
\includegraphics[width=1\textwidth]{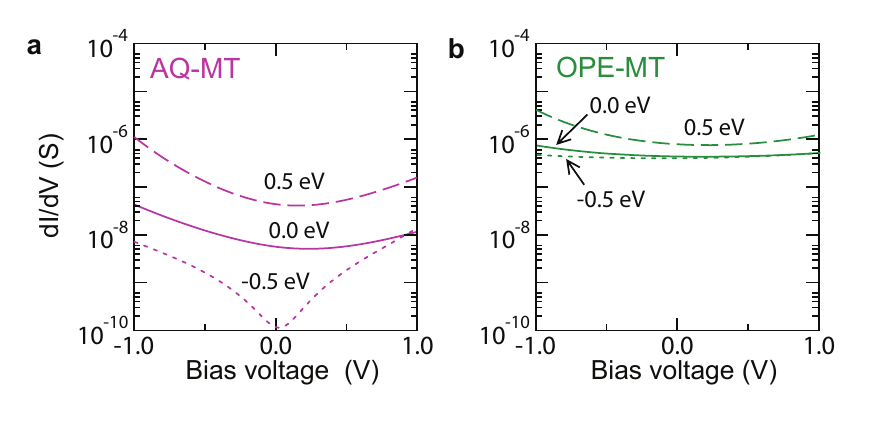}
\caption{\textbf{Calculated $\mathbf{dI/dV}$-curves for AQ-MT and OPE3-MT.}  \textbf{a}, $dI/dV$ for AQ-MT as computed from the transmission function in Fig. 2a.
$T(E)$ was shifted by $\Delta E = 0.0$ eV, $\Delta E = 0.5$ eV and $\Delta E = -0.5$ eV, relative to $E_F$. We include asymmetry in the bias drop through a
parameter $\eta=0.6$ (Supplementary Methods). To account for multiple molecules contacting the AFM-tip, we have multiplied the transmission function by a
factor of 100. For $\Delta E=-0.5\,$eV both the shape and the range of $dI/dV$, spanning two orders of magnitude, are in excellent agreement with the
experiments (Fig. 3b). \textbf{b}, Similar calculation for OPE3-MT. Asymmetric curves and higher conductance values with smaller variation are found,
consistent with Fig. 3d.}
\end{figure}

In summary, our charge transport data provide direct evidence for destructive quantum interference in two-terminal molecular junctions. The
interference effects are intimately linked to the shapes and energies of the molecular orbitals and can thus be controlled by chemical design. The
fact that interference in molecules is present at room temperature opens the road to a new type of molecular devices. Specifically, these include
interference controlled molecular switches with very large on-off ratios\cite{vanderMolen2010,Dijk2006} and novel thermoelectric devices, with thermopower
values tunable in magnitude and sign\cite{Bergfield2009}.\\

* These authors contributed equally to this work

\section{Method Summary}
Samples were prepared by thermal deposition of 5 nm chromium and 200 nm gold onto silicon/silicon oxide substrates. These freshly prepared samples were
immediately transferred into a nitrogen-filled glove-box. The molecular wires were dissolved in dry chloroform (AC-DT, AQ-DT, AQ-MT) or in dry THF (OPE3-DT
and OPE3-MT) at 0.5 mM, in this glove-box. We added 10\% (v/v) degassed triethylamine to these solutions to deprotect the thiol groups and immersed the
gold samples for 2 days, to form densely-packed self-assembled monolayers \cite{Valkenier2011} as confirmed by ellipsometry and XPS (see Supplementary Methods).
After immersion, the samples were rinsed three times with clean chloroform or THF, and dried in the glove-box. The synthesis of AQ-DT and the
characterization of all five molecular wires is reported in the Supplementary Methods. Transport experiments were performed on a Digital Instrument Multimode-AFM
with a Nanoscope III controller. The conductance measurements themselves were controlled externally (see Supplementary Methods). Calculations of junction
geometries and transmission functions were performed with the GPAW density functional theory code using an atomic orbital basis set corresponding to
double-zeta plus polarization and the Perdew-Burke-Ernzerhof exchange-correlation functional. Before calculating the transmission functions, the occupied
and unoccupied molecular orbitals were shifted in energy in order to account for self-interaction errors and missing image charge effects in the DFT
description. This approach (DFT + $\Sigma$) was recently found to systematically improve the DFT-conductance values
\cite{Quek2009} (see Supplementary Methods).\\

\textbf{Acknowledgements} We are grateful to Tjerk Oosterkamp and Federica Galli for making their equipment and expertise available to us. We thank Jan van Ruitenbeek, Marius
Trouwborst for discussions and Daniel Myles for his initial synthetic efforts. This study was financed by a VIDI-grant (SJvdM) of the Netherlands
Organization for Scientific Research (NWO) as well as by the Dutch Ministry of Economic Affairs via NanoNed (HV, project GMM.6973).\\

\textbf{Author Contributions} CMG and SJvdM performed the AFM measurements and the data analysis; HV and JCH designed and synthesized the molecules, made and characterized the SAM's; TM and KST performed the calculations; CG, HV, JCH and SJvdM designed the experiment. All the authors discussed the results and commented on the manuscript.\\

\textbf{Author Information} The authors declare no competing financial interests.
Correspondence should be addressed to SJvdM (Molen@physics.leidenuniv.nl)
\bibliography{literature_QI}
\bibliographystyle{naturemag}
\clearpage

\end{document}